\documentclass[12pt,a4]{article}
\usepackage{latexsym,epsfig,amssymb,euscript,amsmath,verbatim,cite, subfigure,lmodern,color}
\usepackage{float}
\usepackage{verbatim} 
\newcommand{\be}{\begin{equation}}
\newcommand{\ee}{\end{equation}}
\newcommand{\bea}{\begin{eqnarray}}
\newcommand{\eea}{\end{eqnarray}}

\numberwithin{equation}{section}
\usepackage{bm}
\usepackage{hyperref}

\begin{document}
\pagestyle{empty}
\vspace{1.8cm}

\begin{center}
{\LARGE{\bf  {Analytic DC thermo-electric conductivities in holography with massive gravitons}}}

\vspace{1cm}

{\large{Andrea Amoretti$^{a,}$\footnote{\tt andrea.amoretti@ge.infn.it },
Alessandro Braggio$^{a,b,}$\footnote{\tt alessandro.braggio@spin.cnr.it }, 
Nicola Maggiore$^{a,}$\footnote{\tt nicola.maggiore@ge.infn.it },\\ 
Nicodemo Magnoli$^{a,}$\footnote{\tt nicodemo.magnoli@ge.infn.it},
Daniele Musso$^{c,}$\footnote{\tt dmusso@ictp.it} 
\\[1cm]}}

{\small{
{}$^a$  Dipartimento di Fisica, Universit\`a di Genova,\\
via Dodecaneso 33, I-16146, Genova, Italy\\and\\I.N.F.N. - Sezione di Genova\\
\medskip
{}$^b$  CNR-SPIN, Via Dodecaneso 33, 16146, Genova, Italy\\
\medskip
{}$^c$ Abdus Salam International Centre for Theoretical Physics (ICTP)\\
Strada Costiera 11, I 34014 Trieste, Italy
}}
\vspace{1cm}

{\bf Abstract}

We provide an analytical derivation of the thermo-electric transport 
coefficients of the simplest momentum-dissipating model in gauge/gravity 
where the lack of momentum conservation is realized by means of explicit 
graviton mass in the bulk. We rely on the procedure recently described
by Donos and Gauntlett for
holographic models where momentum dissipation is realized through non-trivial scalars.
The analytical approach confirms and supports the results found previously by means 
of numerical computations and the associated holographic renormalization procedure.
Importantly, it also provides a precise identification of the range of validity of the hydrodynamic 
approximation.

\end{center}

\newpage

\setcounter{page}{1} \pagestyle{plain} \renewcommand{\thefootnote}{\arabic{footnote}} \setcounter{footnote}{0}

\tableofcontents

\section{Introduction and conventions}

The purpose of the present letter is to provide an analytic derivation of the DC thermo-electric transport coefficients
for holographic models with massive gravitons in the bulk. These formul\ae\ were recently proposed on the basis
of numerical computations in \cite{Amoretti:2014zha}. Holographic models featuring momentum dissipation due to 
specific scalar field setups have been treated analytically in \cite{Donos:2014cya}. As suggested by the authors 
of \cite{Donos:2014cya}, the same analytical approach can also be exported to holographic models where the 
momentum dissipation is simply realized by means of explicit mass terms for the bulk gravitons. In line with 
this hint, we apply the method of \cite{Donos:2014cya} to provide an analytic derivation of the formul\ae\
emerged from the numerical analysis of the thermo-electric transport of the holographic massive gravity model proposed
in \cite{Vegh:2013sk}.

The approach described in \cite{Donos:2014cya} to compute the DC thermo-electric response is 
based on the analysis of quantities which do not evolve from the IR to the UV. In other words, it
relies on a ``membrane paradigm'' \cite{Iqbal:2008by} for massive gravity bulk models.
The analytic results corroborate our previous numerical results about the behaviour of the DC transport coefficients
which were inspired by field theoretical expectation for systems featuring elastic scattering due to 
impurities \cite{sachdev}.

In order to fix the conventions, let us recall the main properties of the massive gravity model which we have analyzed 
in \cite{Amoretti:2014zha} (previously studied in \cite{Vegh:2013sk,Davison:2013jba,Blake:2013bqa})%
\footnote{For the details of the computations and the physical meaning of this model we refer to 
\cite{Amoretti:2014zha,Vegh:2013sk,Davison:2013jba,Blake:2013bqa}.}. The gravitational action of the model is 
\begin{equation}
\label{massivelag}
\begin{split}
S = \int d^4x\ & \sqrt{-g} \left[ \frac{1}{2 \kappa_4^2}\left(R+\frac{6}{L^2}+\beta \left([\mathcal{K}]^2
-[\mathcal{K}^2]\right) \right)-\frac{1}{4 q^2}F_{\mu \nu}F^{\mu \nu} \right]\\
&+\frac{1}{2 \kappa_4^2} \int_{z=z_{UV}} d^3 x\ \sqrt{-g_b} \  2 K \ + \frac{1}{2 \kappa_4^2} \int_{z=z_{UV}} d^3 x\ \sqrt{-g_b}\ \frac{4}{L} \ ,
\end{split}
\end{equation}
where $\beta$ is an arbitrary parameter having the dimension of a mass squared%
\footnote{We consider just the terms which are quadratic in $\mathcal{K}$. Linear terms or higher-power
terms in $\mathcal{K}$ could be considered as well, see \cite{Vegh:2013sk}. The motivation for retaining just 
the quadratic terms in $\mathcal{K}$ is twofold: the proven absence of ghosts and the fact that
the bulk gauge field has the same UV fall-off's as in the standard massless gravity case.} and, as usual,
$L$ represents the curvature radius of the asymptotically $AdS$ bulk solutions.
The matrices $\mathcal{K}^{\mu}_{\; \nu}$ and $\left(\mathcal{K}^2\right)^{\mu}_{\; \nu}$ are defined in the following way:
\begin{equation}
\begin{split}
&(\mathcal{K}^2)^\mu_{\ \nu}\equiv g^{\mu \rho}f_{\rho \nu} \ , \qquad \mathcal{K} \equiv \left( \sqrt{ \mathcal{K}^2} \right)^{\; \mu}_{\; \; \;\nu}\ , \text{ with } f_{\mu \nu}= \text{diag}(0,0,1,1) \ .
\end{split}
\end{equation}
The small square brackets  in \eqref{massivelag} represent the trace operation
and $f_{\mu\nu}$ is a non-dynamical fiducial metric which explicitly breaks the bulk 
diffeomorphisms along the spatial $x,y$ directions%
\footnote{For more details look at \cite{Vegh:2013sk} and references therein.}.
In \eqref{massivelag} we have also added to the bulk action two boundary terms. 
The first one is the usual Gibbons-Hawking term, expressed in terms of the induced metric $(g_b)_{\mu \nu}$ 
and the trace $K$ of the extrinsic curvature $K_{\mu \nu}$ on the manifold $z=z_{UV}$%
\footnote{Actually $z_{UV}$ represents a UV cut-off that is eventually sent to zero 
in the final step of the holographic renormalization procedure.}, 
which is necessary to have a well-defined bulk variational problem; 
the second one is a counter-term necessary in order to make the on-shell action finite. 
This model admits black-brane solutions corresponding to the following radial ansatz
\begin{equation}\label{bbm}
\begin{split}
&ds^2=\frac{L^2}{z^2} \left[-f(z) dt^2 + dx^2 + dy^2 + \frac{1}{f(z)} dz^2\right]\ , \qquad A=\phi(z)\, dt\ ,\\
&\phi(z)= \mu - q^2 \rho z = \mu \left(1-\frac{z}{z_h} \right) \ , \qquad \rho \equiv \frac{\mu}{q^2 z_h} \ , \\
&f(z)= 1 - \left(1 + \frac{z_h^2 \gamma^2 \mu^2}{2 L^2}\right) \left(\frac{z}{z_h}\right)^3 
 + \frac{z_h^2 \gamma^2 \mu^2}{2 L^2} \left(\frac{z}{z_h}\right)^4 +\beta z_h^2 \left(\frac{z^2}{z_h^2}-\frac{z^3}{z_h^3}\right)\ ,
\end{split}
\end{equation}
where $\gamma = \kappa_4/q$ and $z_h$ is the horizon radius defined by $f(z_h)=0$.
In \eqref{bbm} $\mu$ is the chemical potential associated to the charge density $\rho$. 
The other thermodynamical quantities, namely the temperature $T$, the energy density $\mathcal{E}$, 
the pressure $P$ and the entropy density $\mathcal{S}$ are (see \cite{Amoretti:2014zha})
\begin{equation}
\label{themodyna}
T= - \frac{f'(z_h)}{4 \pi}  =-\frac{\gamma ^2 \mu ^2 z_h}{8 \pi  L^2} +\frac{3}{4 \pi z_h}+\frac{\beta  z_h}{4 \pi } \ ,
\end{equation}
\begin{equation}
\label{thermodyna1}
\mathcal{S}=\frac{2 \pi}{\kappa_4^2}\frac{L^2}{z_h^2} \ , \ \ \ \ \
\mathcal{E}=\frac{L^2}{z_h^3 \kappa_4^2}  + \frac{\mu^2}{2 q^2z_h}+ \frac{L^2 \beta}{z_h \kappa_4^2} \ , \ \ \ \ \
P=\frac{L^2}{2 \kappa_4 ^2 z_h^3}+\frac{\mu ^2}{4 q^2 z_h}-\frac{\beta  L^2}{2 \kappa_4 ^2 z_h} \ .
\end{equation}

We underline that holographic massive gravity models as well as 
other specific setups (e.g. \cite{Donos:2012js,Donos:2013eha}) realize momentum dissipation while preserving the 
homogeneity of the bulk differential problem. This leads to the noteworthy technical advantage of
dealing with ordinary differential equations instead of partial differential equations.
\section{DC transport coefficients computation}
\subsection{The electric conductivity and the Seebeck coefficient}
\label{sec1}

Due to the isotropy of the system we are allowed to consider just perturbations 
in the $x$ direction without loss of generality. Then the static electric conductivity 
$\sigma_{DC}$ and the Peltier coefficient $\Pi_{DC}$ are defined in terms of the electric field $E_x$, 
the charge density current $J_x$ and the heat flow $Q_x$ in the following way
\begin{equation}
\label{sigmaalpha}
\sigma_{DC} \equiv \left. \frac{J_x}{E_x} \right|_{\nabla_x T=0} \ , \qquad \Pi_{DC}\equiv  \left. \frac{Q_x}{E_x} \right|_{\nabla_x T=0} \ .
\end{equation}
The definitions \eqref{sigmaalpha} imply that, in order to compute $\sigma_{DC}$
and $s_{DC}$, we must consider non-zero electric field and vanishing thermal gradient.

Inspired by \cite{Donos:2014cya}, we turn on the following fluctuating components
\begin{eqnarray}\label{ansaflu}
 a_x(t,z) &=& - E\, t + \tilde{a}_x(z)\ ,\\
 h_{tx}(t,z) &=& \tilde{h}_{tx}(z)\ ,\\
 \label{ansaflu1}
 h_{zx}(t,z) &=& \tilde{h}_{zx}(z)\ ,
\end{eqnarray}
where the temporal dependence of the 4-potential $a_\mu$ corresponds to a constant electric field
$E$ along $x$ and with $h$ we denote fluctuations of the metric components.

The set of coupled linearized equations of motion for the fluctuating fields are
\begin{equation}
\label{eq1}
  \tilde{h}_{tx}''(z)
 +\frac{2}{z}\, \tilde{h}_{tx}'(z)
 + 2 \left(\frac{\beta}{f(z)}
 -\frac{1}{z^2} \right) \tilde{h}_{tx}(z)
 -\frac{2 \gamma^2 \mu}{z_h}\, \tilde{a}_x'(z) = 0\ ,
\end{equation}

\begin{equation}\label{hzx}
 \frac{E\, z^2 \mu\, \gamma^2}{z_h f(z)}
 + \left(\frac{z^4 \mu^2 \gamma^2}{2 z_h^2 L^2}
 - z f'(z)
 + 3f(z)
 - 3 \right) \tilde{h}_{zx}(z) = 0\ ,
\end{equation}

\begin{equation}
\label{eq2}
 \tilde{a}_x''(z)
+\frac{f'(z)}{f(z)} \tilde{a}_x'(z)
-\frac{z^2 \mu }{L^2 z_h f(z)} \tilde{h}_{tx}'(z)
-\frac{2 z \mu }{L^2 z_h f(z)} \tilde{h}_{tx}(z) = 0\ .
\end{equation}
Note that equation \eqref{hzx} for $\tilde{h}_{zx}$ can be solved algebraically and,
recalling the explicit expression of the emblackening factor $f(z)$ given in
\eqref{bbm}, the solution can be expressed as follows
\begin{equation}
 \tilde{h}_{zx}(z) = - \frac{E \gamma^2 \mu}{z_h \beta f(z)}\ .
\end{equation}

In order to completely determine the solution of the remaining two equations \eqref{eq1} and \eqref{eq2},
we have to provide suitable boundary conditions for the fluctuation fields $h_{tx}(t,z)$ and $a_x(t,z)$
at the conformal boundary $z=0$ and at the horizon $z=z_h$.
At the horizon we require the regularity of the fluctuations; this requirement can be easily fulfilled
by switching to the Eddington-Finkelstein time coordinate
\begin{equation}
 v = t - \frac{1}{4 \pi T} \log\left(\frac{z_h-z}{L}\right)\ ,
\end{equation}
leaving untouched all the other coordinates. 
From the IR regularity requirement for all the metric components in the new coordinate system we
derive the behavior of $\tilde{h}_{tx}$ at the horizon, namely
\begin{equation}\label{nhhtx}
 \tilde{h}_{tx}(z) = - \frac{E \gamma^2 \mu}{z_h \beta} + \mathcal{O} (z-z_h)\ .
\end{equation}
An analogous regularity requirement at the horizon applied to the gauge field yields
\begin{equation}\label{nha}
\tilde{a}_x(z) = \frac{E}{4 \pi T} \log\left(\frac{z_h-z}{L}\right) + \mathcal{O}(z-z_h) \ .
\end{equation}

Considering the conformal boundary located at $z=0$, we have to furnish boundary conditions in such 
a way that the dual system has an external electric field and vanishing thermal gradient. 
Given the ansatz \eqref{ansaflu}, the small-$z$ leading behavior of the fluctuating field 
$a(t,z)$ corresponds to a constant electric field $E$. According to the standard holographic 
dictionary, the coefficient of the subleading fall-off in $z$ of the 
field $a(t,z)$ corresponds to the charge density current $J_x$; namely $\tilde{a}_x \sim J_x z$
at small $z$. Regarding the metric fluctuation $\tilde{h}_{tx}$, from equation \eqref{eq1} it is easy to see that 
it has two independent behaviors in the near-boundary region, i.e. $z$ and $z^{-2}$. The holographic dictionary prescribes that
imposing that there are no sources associated to thermal gradients corresponds to setting 
to zero the coefficient of the leading $z^{-2}$ term. All in all, the set of boundary conditions 
that we have just illustrated determines the solution of the differential equations \eqref{eq1} and \eqref{eq2}
completely. 

In order to compute the DC transport coefficients it is fundamental to note that there are two linear combinations 
of the fluctuations which are independent of the coordinate $z$ and are
respectively related to the charge density current and the heat current. The first conserved current is
\begin{equation}\label{curre}
 \bar{J}^\mu = -\frac{\sqrt{-g}}{q^2} F^{z\mu} \ ,
\end{equation}
with $\mu= t, x, y$. Given the ansatz \eqref{ansaflu}, the only non-zero component of the Maxwell equation 
$\sqrt{-g}\, \nabla_M F^{M N} = \partial_z (\sqrt{-g} F^{z x}) = 0$ states that $\bar J^x$ is independent of the radial coordinate $z$
and it assumes the following explicit form in terms of the fluctuating fields
\begin{equation}
\label{conserv1}
 \bar{J}^x = -\frac{\sqrt{-g}}{q^2} F^{zx} 
     = \frac{z^2 \mu}{L^2 q^2 z_h}\, \tilde{h}_{tx}(z) 
       -\frac{f(z)}{q^2}\, \tilde{a}_x'(z) \ .
\end{equation}
Relying on the ``radial conservation'' we can compute $\bar{J}^x$ both at the horizon $z=z_h$ and at the boundary $z=0$
knowing that the two results must correspond. Computing it at $z=0$ and recalling 
the above-mentioned UV behaviors,
it is possible to see that this quantity is actually the charge density current $J^x$ of the dual field theory. 
Then, evaluating \eqref{conserv1} at $z=z_h$ we find: 
\begin{equation}
J^x= \left( \frac{1}{q^2} - \frac{\gamma^2 \mu^2}{L^2 q^2 \beta} \right) E \ .
\end{equation}
The electric conductivity $\sigma_{DC}$ is now easily computed by means of \eqref{sigmaalpha},
\begin{equation}\label{elcon}
 \sigma_{DC} = \frac{J^x}{E} = \frac{1}{q^2} - \frac{\gamma^2 \mu^2}{L^2 q^2 \beta}\ ,
\end{equation}
which corresponds exactly with the analytical expression found in \cite{Blake:2013bqa}.

The second conserved quantity, which is related to the heat current, is subtler to identify. Indeed, as noted in \cite{Donos:2014cya}, it is associated to the existence of the Killing vector $k=\partial_t$ and it assumes the following form:

\begin{eqnarray}
\nonumber
\bar{Q}&=&\frac{\sqrt{-g}}{\kappa_4^2}\nabla^z k^x-\phi J^x=\\ 
\label{conserv2}
&& \left(\frac{\phi \  \tilde{a}_x'}{q^2}+\frac{\tilde{h}_{tx}'}{2 \kappa_4 ^2}+\frac{\tilde{h}_{tx}}{\kappa_4 ^2 z}\right)f+
   \left(-\frac{f'}{2 \kappa_4 ^2}-\frac{\mu  z^2 \phi }{L^2 q^2 z_h}\right)\tilde{h}_{tx} \ ,
\end{eqnarray}
where in the second passage we gave an explicit expression of $\bar Q$ in terms of the fluctuation fields.
Evaluating $\bar Q$ on the equations of motion \eqref{eq1} and \eqref{eq2} and keeping into account the particular 
form of the background quantities $f$ and $\phi$  \eqref{bbm}, one has that $\partial_z \bar{Q}=0$.
Namely $\bar Q$ is radially conserved.

Having obtained a radially conserved quantity, in order to repeat the same steps done previously for the electric conductivity, we have to prove that $\bar{Q}$ coincides, if evaluated on the boundary, with the heat current in the $x$ direction $Q^x=T^{tx}-\mu J^x$ of the dual field theory. Actually, in this case, the proof of this statement is straightforward. In fact, $J^x$ is constant along the radial direction and $\phi(0)=\mu$, and consequently the term $\phi \, J^x$ reduces to $\mu\, J^x$
when evaluated at the conformal boundary. Moreover, it is not difficult to see that $\frac{\sqrt{-g}}{\kappa_4^2}\nabla^z k^x$ 
coincides with the linearized $tx$ component of the stress-energy tensor of the dual field theory%
\footnote{To have an explicit expression of the stress-energy tensor we refer for instance to \cite{Balasubramanian:1999re}}
evaluated at the conformal boundary $z=0$, namely 
\begin{equation}
T^{tx}=\frac{L^5}{\kappa_4^2 z^5} \left( -K^{tx}+K g_b^{tx}+\frac{2}{L}g_b^{tx} \right)=
\frac{\tilde{h}_{tx}'}{2 \kappa_4 ^2 \sqrt{f}}-\frac{\tilde{h}_{tx}}{\kappa_4 ^2 z \sqrt{f}}+\frac{2 \tilde{h}_{tx}}{\kappa_4 ^2 z f}\ .
\end{equation} 

Finally, having associated the quantity $\bar Q$ with the heat flow, the Peltier coefficient 
is straightforwardly obtained evaluating $\bar{Q}$ at the horizon $z=z_h$ and relying 
on the IR behavior of the fluctuating fields. We obtain
\begin{equation}
\label{seebeck}
\Pi_{DC}= \frac{\bar{Q}}{E}=-\frac{2 \pi  \mu }{\beta  q^2 z_h} \left( -\frac{\gamma ^2 \mu ^2 z_h}{8 \pi  L^2} +\frac{3}{4 \pi z_h}+\frac{\beta  z_h}{4 \pi } \right) \ .
\end{equation} 

The expressions for the DC electrical conductivity \eqref{elcon} and the Peltier coefficient 
\eqref{seebeck} obtained here with an analytical computation along the lines described by \cite{Donos:2014cya}
coincide with those found in \cite{Amoretti:2014zha} using numerical methods. Note that the term inside the parenthesis of equation \eqref{seebeck} coincides exactly  with the temperature \eqref{themodyna}. Consequently, by Onsager reciprocity (see later), one has $s_{DC}=\frac{\Pi_{DC}}{T}$, finding exactly formula (1.4) of \cite{Amoretti:2014zha}, which was satisfied by numerical data.

\subsection{Thermal conductivity and Onsager reciprocity}

The thermal conductivity $\bar{\kappa}_{DC}$ is defined as 
\begin{equation}
\bar{\kappa}_{DC}\equiv \left. \frac{Q_x}{-\nabla_x T} \right|_{E_x=0} \ .
\end{equation}
Symmetrically to what we have done in the previous Section, to compute this quantity
we must consider a thermal gradient at vanishing electric field. To this end, we rely on the elegant 
method described in \cite{Donos:2014cya} (to which we refer the reader for further details); 
we consider the following set of fluctuations
\begin{eqnarray}\label{ansatzflu1}
 a_x(t,z) &=& \alpha_2\, \phi(z)\, t + \tilde{a}_x(z)\ ,\\ \label{ansatzflu11}
 h_{tx}(t,z) &=&-\alpha_2\, \frac{L^2}{z^2}\, f(z) \, t+ \tilde{h}_{tx}(z)\ ,\\ \label{ansatzflu111}
 h_{zx}(t,z) &=& \tilde{h}_{zx}(z)\ .
\end{eqnarray}
Note that, following the standard holographic prescription (see for instance \cite{Hartnoll:2009sz,Herzog:2009xv,Musso:2014efa}), 
the coefficient $\alpha_2$ corresponds to the thermal gradient $-\nabla_x T/T$. It actually represents the 
thermal source dual to the boundary value of the $h_{tx}$ bulk field (see \eqref{ansatzflu11}).

Considering the ansatz \eqref{ansatzflu1}, \eqref{ansatzflu11} and \eqref{ansatzflu111}, one finds that the equations 
of motion for $\tilde{h}_{tx}$ and $\tilde{a}_x$ are the same as those given in Section \ref{sec1}, namely \eqref{eq1} and \eqref{eq2}. 
The equation for $\tilde{h}_{zx}$ looks slightly different but is still algebraically solvable; its solution being
\begin{equation}
\tilde{h}_{zx}(z)=-\frac{L^2 \alpha_2 \left\{2  L^2 q^2 \left[3 z \left(\beta  z_h^2+1\right)-2 \beta  z_h^3\right]-
   \kappa_4 ^2 \mu ^2 z z_h^2\right\}}{2 \beta  z (z-z_h) \left\{\kappa_4 ^2 \mu ^2 z^3 z_h-2 L^2 q^2 \left[z^2 \left(\beta 
   z_h^2+1\right)+z z_h+z_h^2\right]\right\}}\ .
\end{equation}
As regards the boundary conditions, the regularity of the fluctuations at the horizon, in this case, implies that
\begin{equation}
\label{boundary21}
\tilde{h}_{tx}(z)=-\frac{\alpha_2 \left(2  L^2 q^2 \left(\beta  z_h^2+3\right)- \kappa_4 ^2 \mu ^2
  z_h^2\right)}{4 \beta  z_h^3 q^2}
+\frac{\alpha_2 L^2 f(z) }{2 \pi  T z^2}\log \left(\frac{z_h-z}{L}\right) +\mathcal{O}(z-z_h) \ ,
\end{equation}
and
\begin{equation}
\label{boundary22}
\tilde{a}_x=\mathcal{O}(z-z_h) \ .
\end{equation}
Moreover, at the boundary $z=0$ we require that $\tilde{h}_{tx}$ is proportional to $z$ 
and, as before, we have $\tilde{a}_x = J_x z + ...$, where the ellipsis indicates higher power of $z$.

Importantly, since the equations of motion for $\tilde{h}_{tx}$ and $\tilde{a}_x$ are the same as in
the case studied in the previous Section, the quantities $\bar{J}^x$ \eqref{conserv1} and $\bar{Q}$ 
\eqref{conserv2} are still independent of $z$. 
Enforcing the near-boundary conditions discussed previously, we have that also in this case $\bar{J}^x$ 
corresponds to the $x$ component of the dual charge density current. However, the relation between $\bar{Q}$ 
and the heat current $Q^x$ is in this case more subtle than before. Indeed the $tx$ component of the holographic 
stress-energy tensor evaluated on the ansatz \eqref{ansatzflu1}, \eqref{ansatzflu11} and \eqref{ansatzflu111} 
for the fluctuations assumes the following form 
\begin{equation}
\label{stress1}
T^{tx}=t \left(-\frac{\alpha_2 L^2 f'}{2 \kappa_4 ^2 z^2 \sqrt{f}}+\frac{2 \alpha_2 L^2 \sqrt{f}}{\kappa_4 ^2
   z^3}-\frac{2\alpha_2 L^2}{\kappa_4 ^2 z^3}\right)+\frac{\tilde{h}_{tx}'}{\kappa_4 ^2 \sqrt{f}}-\frac{\tilde{h}_{tx}}{\kappa_4 ^2
   z \sqrt{f}}+\frac{2 \tilde{h}_{tx}}{\kappa_4 ^2 z f}\ .
\end{equation}
Therefore the quantity $\frac{\sqrt{-g}}{\kappa_4^2}\nabla^z k^x$ (contained in $\bar Q$)
computed at $z=0$ equals just the time-independent part of $T^{tx}$ (i.e. the last three terms in \eqref{stress1}). 
Nevertheless, as discussed in \cite{Donos:2014cya}, in order to compute the DC response, only the time-independent 
part of $T^{tx}$ is relevant. To the purpose of computing the DC thermal conductivity we can then assume that $\bar{Q}$ 
corresponds to the heat current, and evaluating it at the horizon, and considering in this case the horizon conditions \eqref{boundary21} and \eqref{boundary22}, we find
\begin{equation}
\bar{\kappa}_{DC} =\frac{1}{T}\frac{\bar{Q}}{\alpha_2}
=\frac{\pi  \left[\gamma ^2 \mu ^2 z_h^2-2 L^2 \left(\beta  z_h^2+3\right)\right]}{2 \beta  \kappa_4 ^2 z_h^3} \ ,
\end{equation} 
which agrees with our previous numerical result reported in \cite{Amoretti:2014zha}.

Eventually, relying on the boundary conditions \eqref{boundary21} and \eqref{boundary22} we evaluate the charge density current $\bar{J}^x$ \eqref{conserv1} at $z=z_h$.
We can then prove that the Onsager reciprocity relation holds, namely
\begin{equation}
s_{DC}=  \left. \frac{J_x}{- \nabla_x T} \right|_{E_x =0}=-\frac{2 \pi  \mu }{\beta  q^2 z_h}=\frac{\Pi_{DC}}{T} \ ,
\end{equation}
where the last identity can be obtained by using \eqref{seebeck} and dividing it by the temperature \eqref{themodyna}.\\
Said otherwise, the conductivity matrix is symmetrical and the Peltier coefficient $\Pi_{DC}$ is equal to the Seebeck coefficient multiplied by $T$.

\section{Discussion and conclusions}

In the present letter we have computed analytically the DC thermo-electric transport coefficients 
for holographic models featuring momentum dissipation realized by explicit graviton mass in the bulk.
We followed precisely the method illustrated in \cite{Donos:2014cya} and applied it to the holographic model 
first introduced in \cite{Vegh:2013sk}. Introducing the explicit form of the dissipation rate for the model
\begin{equation}
\label{taumassive1}
\tau^{-1}=-\frac{\mathcal{S} \beta}{2 \pi (\mathcal{E}+P)} \ , 
\end{equation}
which was found in \cite{Davison:2013jba} by studying the pole of the propagators in the hydrodynamic regime, and using the explicit expression of the thermodynamical quantities \eqref{themodyna} and \eqref{thermodyna1},  the transport coefficients assumes the following beautiful form:
\begin{equation}
\label{coeffintrobal}
\begin{split}
\sigma_{DC}  =  \frac{1}{q^2}+\frac{\rho^2}{\mathcal{E}+P}\tau \ ,  \qquad s_{DC}  =   \frac{\mathcal{S} \rho}{\mathcal{E}+P}\tau \ ,
\ \ \ \ \ \
\bar{\kappa}_{DC}  =  \frac{\mathcal{S}^2 T}{\mathcal{E}+P}\tau \ ,
\end{split}
\end{equation}
Such explicit expressions coincide with those found in \cite{Amoretti:2014zha} relying on a 
numerical analysis. 

The analytical results outlined in this paper put a firmly ground to the  statements made in \cite{Amoretti:2014zha}, which were mainly obtained on purely phenomenological observation.
First of all, we are now able to establish precise limits on the validity of the hydrodynamic regime. In fact in \cite{Amoretti:2014zha} we 
have found that for certain values of the thermodynamic quantities and of the parameters of the model the thermo-electric transport coefficients 
computed in massive gravity are compatible with the hydrodynamic prediction made in \cite{Hartnoll:2007ih}, namely:
\begin{equation}
\begin{split}
\label{coeffintro}
\sigma_{DC}  =  \frac{1}{q^2}+&\frac{\rho^2}{\mathcal{E}+P}\tau \ ,  \qquad s_{DC}  =   -\frac{1}{q^2} \frac{\mu}{T}+\frac{\mathcal{S} \rho}{\mathcal{E}+P}\tau \ ,\\
&\bar{\kappa}_{DC}  =   \frac{1}{q^2}\frac{\mu^2}{T}+\frac{\mathcal{S}^2 T}{\mathcal{E}+P}\tau \ .
\end{split}
\end{equation}
By the comparison of the previous formul\ae \ with the exact results \eqref{coeffintrobal} it is easy to see that, 
in order for the hydrodynamic approximation to be valid, the following inequality must hold:
\begin{equation}\label{fc}
\tau^{-1} \ll \min \left\{  \frac{\mathcal{S}^2 q^2}{\mathcal{E}+P}\frac{T^2}{\mu^2} \ , \ \frac{\mathcal{S} \rho q^2}{\mathcal{E}+P} \frac{T}{\mu} \right\} \ . 
\end{equation} 
Since both the  terms $\frac{\mathcal{S} \rho q^2}{\mathcal{E}+P}$ and $\frac{\mathcal{S}^2 q^2}{\mathcal{E}+P}$ are approximatively constant at sufficiently low $T$, one can see that the first of the two bounds to the dissipation rate $\tau^{-1}$ in \eqref{fc} is the lowest bound near $T/\mu \sim 0$; in contrast the second bound is the most stringent at higher values of the ratio $T/\mu$.

It is also interesting to note that the present analysis allows us to pinpoint a technical aspect of \cite{Amoretti:2014zha}. 
In fact, the analysis described in \cite{Amoretti:2014zha} highlighted
the importance of a careful holographic renormalization procedure to the aim of obtaining 
a physically meaningful thermo-electric response. Specifically, the reduced amount of symmetry 
due to the mass term for the bulk graviton allows for a wider set of possible boundary terms.
Requiring a well-behaved high-frequency response and the absence of unphysical dissipationless 
heat transport fixes the coefficients of the finite counter-terms. This latter feature corresponds 
to the physical need of absorbing a DC delta function in the real part of the thermal conductivity signalled 
by a pole in the corresponding imaginary part. To be more precise, in order to have a well behaved imaginary part of the spectral thermal conductivity in the limit $\omega \rightarrow 0$ we found it necessary to add to the action \eqref{massivelag} the finite counter-term
\begin{equation}
\label{fctt}
 S_{\text{c.t.}}^{\text{(fin)}} = - \frac{1}{4}\ \mathcal{E} \int_{z=z_{UV}} d^3x\ \frac{z}{L}\, \sqrt{-g_{b \; tt}}\, g^{tt}g^{xx}\, h_{tx}h_{tx}\ .
\end{equation}
Note that the coefficients in front of this counterterm is unusually state dependent and the reason of the presence of this term in the action requires a more specific analysis, which we postpone to future studies. However, adding these finite counter-terms to the action 
\eqref{massivelag} would have resulted in the presence of additional terms in the $tx$ component 
of the holographic stress-energy tensor $T^{tx}$ and such additional terms, in accordance with the ansatz \eqref{ansaflu}-\eqref{ansaflu1}, result in a linear 
time-dependent term (which is of no importance in computing the DC response).

Finally, as regards the future perspective, they are manifold. All the extensions of the simplest massive gravity holographic 
model could be addressed from the ``membrane paradigm'' standpoint developed in \cite{Blake:2013bqa} and \cite{Donos:2014cya}.
The hope being to extend the analytical reach in the context of holographic systems featuring momentum dissipation.

\section{Acknowledgements}
We thank the support of INFN Scientific Initiative SFT: ``Statistical Field Theory, Low-Dimensional Systems, Integrable Models and Applications'' and FIRB - ``Futuro in Ricerca 2012''- Project HybridNanoDev RBFR1236VV.

\end{document}